\documentclass[11pt]{article}
\usepackage{graphicx}

\usepackage{color}

\usepackage{amsmath, amssymb,amsfonts,longtable,tabularx}
\usepackage{cite}
\usepackage{colortbl}
\usepackage[all]{xy}
\usepackage{epsfig}
\usepackage{subfigure}

\usepackage{setspace}
\usepackage{caption}
\usepackage{pifont}
\usepackage[table]{xcolor}

\definecolor{gray1}{rgb}{0.9,0.9,0.9}
\definecolor{gray2}{rgb}{0.8,0.8,0.8}

\newcommand{\BigO}[1]{O\left(#1\right)} 

\usepackage[hmargin=2cm, vmargin=2cm]{geometry}

\date{}

\usepackage{cite}

\begin{document}

\begin{flushleft}
{\huge
\textbf{The evolution of complex gene regulation \\ by low specificity binding sites}
}
\bigskip
\\
Alexander J.Stewart$^{1}$, 
Joshua B. Plotkin$^{1,2}$
\\
\bigskip
\bigskip
$^1$ Department of Biology, University of Pennsylvania, Philadelphia, PA 19104, USA
\\
$^2$ E-mail: jplotkin@sas.upenn.edu\\
\end{flushleft}

\vspace{1cm}

\begin{abstract} Transcription factor binding sites vary in their specificity,
both within and between species. Binding specificity has a strong impact on
the evolution of gene expression, because it determines how easily regulatory
interactions are gained and lost. Nevertheless, we have a relatively poor
understanding of what evolutionary forces determine the specificity of binding
sites. Here we address this question by studying regulatory modules composed of
multiple binding sites. Using a population-genetic model, we show that more
complex regulatory modules, composed of a greater number of binding sites, must
employ binding sites that are individually less specific, compared to less complex
regulatory modules.  This effect is extremely general, and it hold regardless of
the regulatory logic of a module. We attribute this phenomenon to the inability of
stabilising selection to maintain highly specific sites in large regulatory
modules.
%This effect is most pronounced at the levels of selection and mutation that have
%been empirically measured for binding sites across a diverse range fo species.
Our analysis helps to explain broad empirical trends in the yeast regulatory
network: those genes with a greater number of transcriptional regulators feature
by less specific binding sites, and there is less variance in their specificity,
compared to genes with fewer regulators. Likewise, our results also help to
explain the well-known trend towards lower specificity in the transcription factor
binding sites of higher eukaryotes, which perform complex regulatory tasks,
compared to prokaryotes.
\end{abstract}

\vspace{1cm}

\section*{Introduction}

Tanscriptional regulators integrate signals from genes and the environment to
ensure that the correct patterns of gene expression are maintained in the
cell \cite{Davidson:2002fk,ABC:Eldar,Guet:2002uq,Li:2010ys,Little:2010zr,ABC:She2002,Tirosh:2006vn}.  This can be a complicated task, particularly in higher eukaryotes where
processes such as cell differentiation and complex inter-cellular signalling
occur \cite{Davidson:2002fk,Li:2010ys}.  Generally, the more complex a signal integration task, the more complex
the pattern of gene regulation required, and on the face of it we might expect
more complex gene regulation to be carried out by binding sites of higher 
specificity -- just as we might expect a very complicated machine to use high
precision components.  In this paper we show that, in fact, the opposite is true:
natural selection favors less specific binding sites in more complex regulatory
modules.  

We use an established biophysical model of transcription factor binding
to describe regulatory modules consisting of multiple transcription factor
binding sites with a range of specificities.  For selection of a given strength on
gene expression we determine the average information of binding sites that
participate in a module of a given size, across all possible regulatory logics.
Our analysis predicts a strong, negative relationship between the information
content of the binding sites maintained in a module, and the size of the
module.  Thus, we predict more complex regulation by less specific binding sites.

This simple but counter-intuitive result helps to explain two broad empirical
patterns in the transcription networks within and between species.  Firstly,
regulatory complexity in eukaryotes is greater than in prokaryotes
\cite{Lynch:2006vn,Janga:2008ys}, and this difference is accompanied by a tendency
towards less informative binding sites for eukaryotic transcription factors
\cite{Lassig:2007fk,Stewart:2012zr}.  Secondly, within the yeast transcription
network, we observe that those genes whose expression is more variable across
environmental conditions, and therefore require more complex integration of
environmental signals, tend to have a greater number of transcription factor
binding sites each with
lower specificity, compared to genes whose expression is less variable across
environmental conditions.  This second empirical observation was previously made
in a study by Bilu and Barkai
%, in which they also observed that there is greater ``fuzziness" of a binding
%sites that occur in a large regulatory modules
\cite{Bilu:2005kx}.  Bilu and Barkai suggested their observed might be explained
either by weaker selection on the expression of genes with highly variable
expression, or by a tendency for multiple co-regulating binding sites to
experience compensatory mutations.  However we will show that no assumptions
about the strength of selection on gene expression, or about the epistatic effects
of mutations among binding sites, are necessary to explain these empirical
patterns.  Rather, it is necessary only to consider the impact of stabilizing
selection on the information content of binding sites maintained in a regulatory
module to explain the observed trends.  Our results are quite general and so they
can account for the broad empirical phenomenon that complex regulation tends to be
carried out by low-specificity transcription factor binding sites.

Our paper is structured as follows: We begin by describing a standard biophysical
model for transcription factor binding, which we use to construct fitness
landscapes for regulatory modules consisting of multiple binding sites that are
selected to execute a given regulatory logic.  We analyse the evolution of these
modules in the limit of weak mutation, and we determine whether, at equilibrium
under stabilizing selection, the binding sites that belong to the module are
likely to be functional or non-functional (i.e., whether they are likely to be
bound by their respective transcription factors, or not).  We determine how the
information content of the transcription factors that belong to a module varies
with the module size, for fixed selection strength and population size. To begin
with, we focus on modules consisting of a single pair of transcription factors,
since this case can be understood analytically.  We then employ evolutionary
simulations to explore larger modules.  Our simulations exhaustively explore the
possible regulatory logics for modules of a given size, and for any given module
they explore a wide range of combinations of binding-site information content.
Finally we consider the impact of variation in the co-expression patterns of
transcription factor proteins on the ability of selection to maintain their binding sites
in a regulatory module.  We compare the results of our analysis to empirical data from the
yeast transcription network, focusing on the information content of binding sites within
regulatory modules.%, and the expression variation of the genes they regulate.

\section*{Model and Results}
\subsection*{Biophysical model of transcription factor binding}
We use a long-established biophysical model of transcription factor binding
\cite{Bintu:2005fk,Buchler:2003uq,Chu:2009vg,Gerland:2002fk,Mustonen:2008zr},
which treats a binding site as a sequence of $n$ consecutive nucleotides for which
there is an associated consensus sequence (or set of sequences) that results in a
minimum binding energy.  Any given realization of $n$ consecutive bases can be
characterized by its number of ``mismatches'', i.e. the number of nucleotide
positions at which it differs from the consensus sequence.  In the standard model,
each such mismatch increases the binding energy of the sequence, compared to the
consensus sequence, by an amount $\epsilon$.  The increase in energy per-mismatch
has been empirically measured to fall within the range 1 and 3 $k_{B}T$
\cite{Gerland:2002cv,Lassig:2007fk}.  The probability $\pi_{i}$ that any given $n$
nucleotide sequence is bound by a transcription factor is determined by the number
of mismatches, $i$, the binding energy per-mismatch, $\epsilon$, and the number of
free transcription factor proteins in the cell, $P$, according to the equation:

\begin{equation}
\pi_{i}=\frac{P}{P+\exp[\epsilon i]}.
\end{equation}
\\
We describe the consensus sequence of a binding site by assuming that each of the
$n$ nucleotide positions can be treated as having a degeneracy, $r$, which
quantifies the (average) number of different bases that can appear at each
position and still result in minimum binding energy.  Thus, if $r=1$, minimum
binding energy is achieved only if each of the $n$ nucleotides adopts a single
specific base.  If $r=2$, minimum binding energy can be achieved if each of the
$n$ nucleotides adopts one of two bases, and so on.  The average degeneracy for a
given transcription factor can be calculated from the position-specific weight
matrix (PSWM) of its binding site \cite{Stewart:2012zr}.

Increasing the average degeneracy $r$ of a consensus sequence increases its
``fuzziness'', and therefore lowers the specificity of the site, since a greater
number of different nucleotide sequences result in minimum binding energy.
Similarly reducing the length of the consensus sequence, $n$, also decreases its
specificity, since fewer nucleotides need to be matched to a specific base to
produce a sequence with minimum binding energy.  In order to compare the
specificities of different binding sites with different lengths and average
degeneracies, we follow the approach used in earlier studies and measure the
information content, $I$, of a PSWM \cite{Dhaeseleer:2006fk,Sengupta:2002fk},
which is given by $I=n\log_{2}\left[\frac{4}{r}\right]$.

\subsection*{Mutation and selection in the weak mutation limit}

We use the probability of binding, $\pi_i$ to construct the fitness landscape of a
regulatory module.  As in previous studies \cite{Mustonen:2008zr}, we assume that
fitness is a linear function of the probability that a binding site under
selection is in fact bound.  Thus, for a single binding site with $i$ mismatched
nucleotides, the fitness $w_i$ is given by $w_i=1-s(1-\pi_i)$. (We later generalize
this to modules composed of multiple binding sites). In the case of a single
binding site, when $\pi_i=1$ the site is always bound producing fitness $w_i=1$.
The parameter $s$ quantifies the reduction in fitness that occurs when the binding
site is unbound, so that if $\pi_i=0$, and the site is always unbound, we
assign fitness $w_i=1-s$.

Following the approach used in previous studies, we  carry out our analysis of
binding site evolution at a single target gene in the limit of weak mutation
\cite{Berg:2004uq,Moses:2003fk,Sella:2005fk}.  This regime is realistic because the
per-nucleotide mutation rate in both prokaryotes and eukaryotes is low,
$\mu\sim\BigO{10^{-8}}$, binding sites are typically short, $n\sim\BigO{10}$, and
selection on conserved binding sites is sufficiently strong, with
$Ns\sim\BigO{10}$, where $N$ is the population size.  In the weak-mutation limit,
evolution occurs through a series of selective sweeps, with new mutations arising
only after earlier mutations have either fixed or gone extinct.
We can calculate the equilibrium distribution $F_i$ of binding sites with $i$ mismatched nucleotides (see SI). 
We find that when there is no selection, $s=0$, then $F_i$ is just a binomial distribution
whose mean is determined by the rates of mutations that increase or decrease $i$. 
However, when selection is present, i.e. $s>0$, the equilibrium distribution $F_i$ is bimodal, 
with one peak occurring at values of $i$ for which $\pi_i\sim1$ and a second peak occurring 
at the neutral equilibrium, $i=n\left(1-\frac{r}{4}\right )$ (Fig.~1a).

\begin{figure}[h] \centering \subfigure{\includegraphics[scale=0.2]{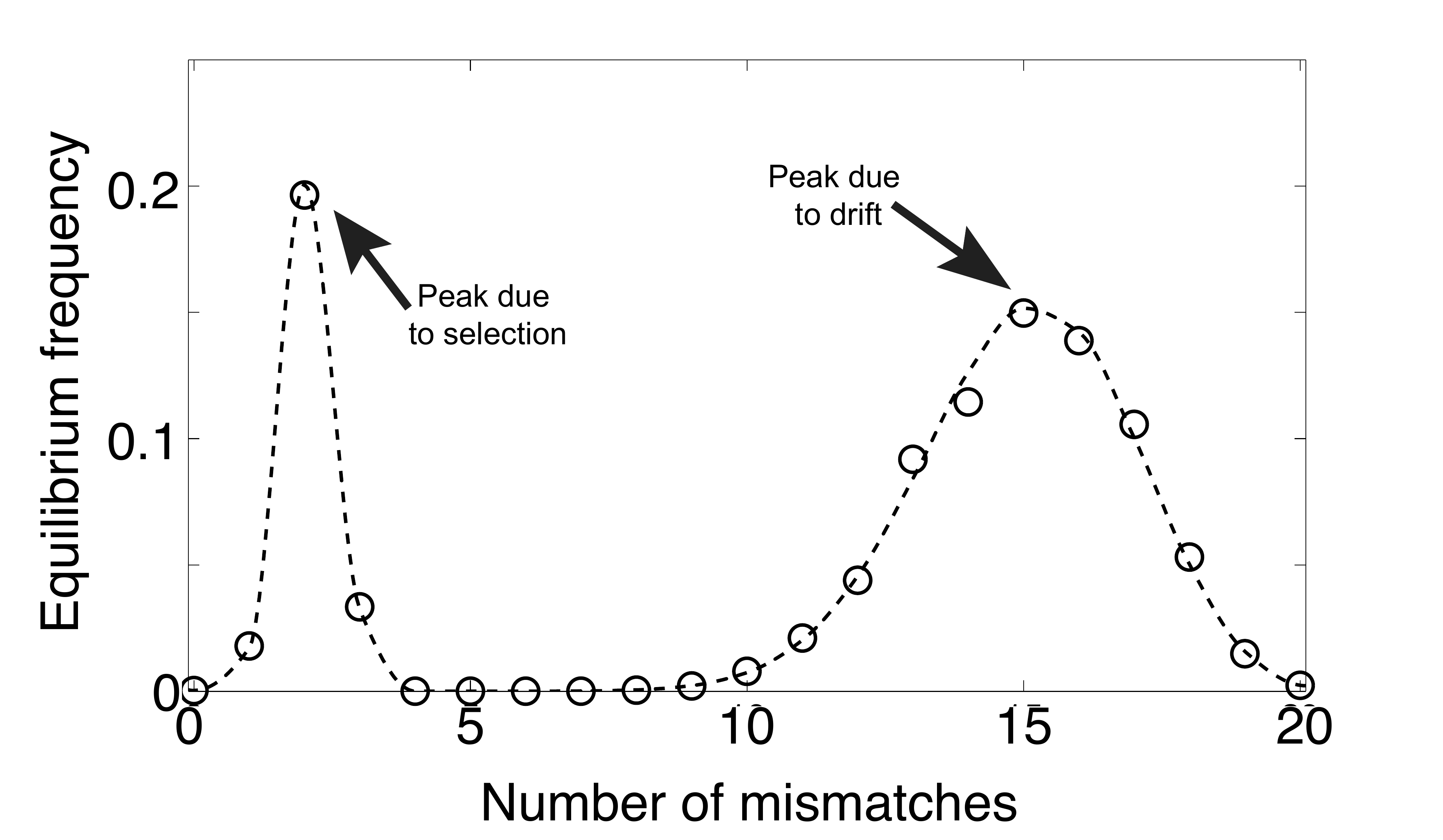}} 
%\subfigure{\includegraphics[scale=0.25]{FunF1}}  
\subfigure{\includegraphics[scale=0.2]{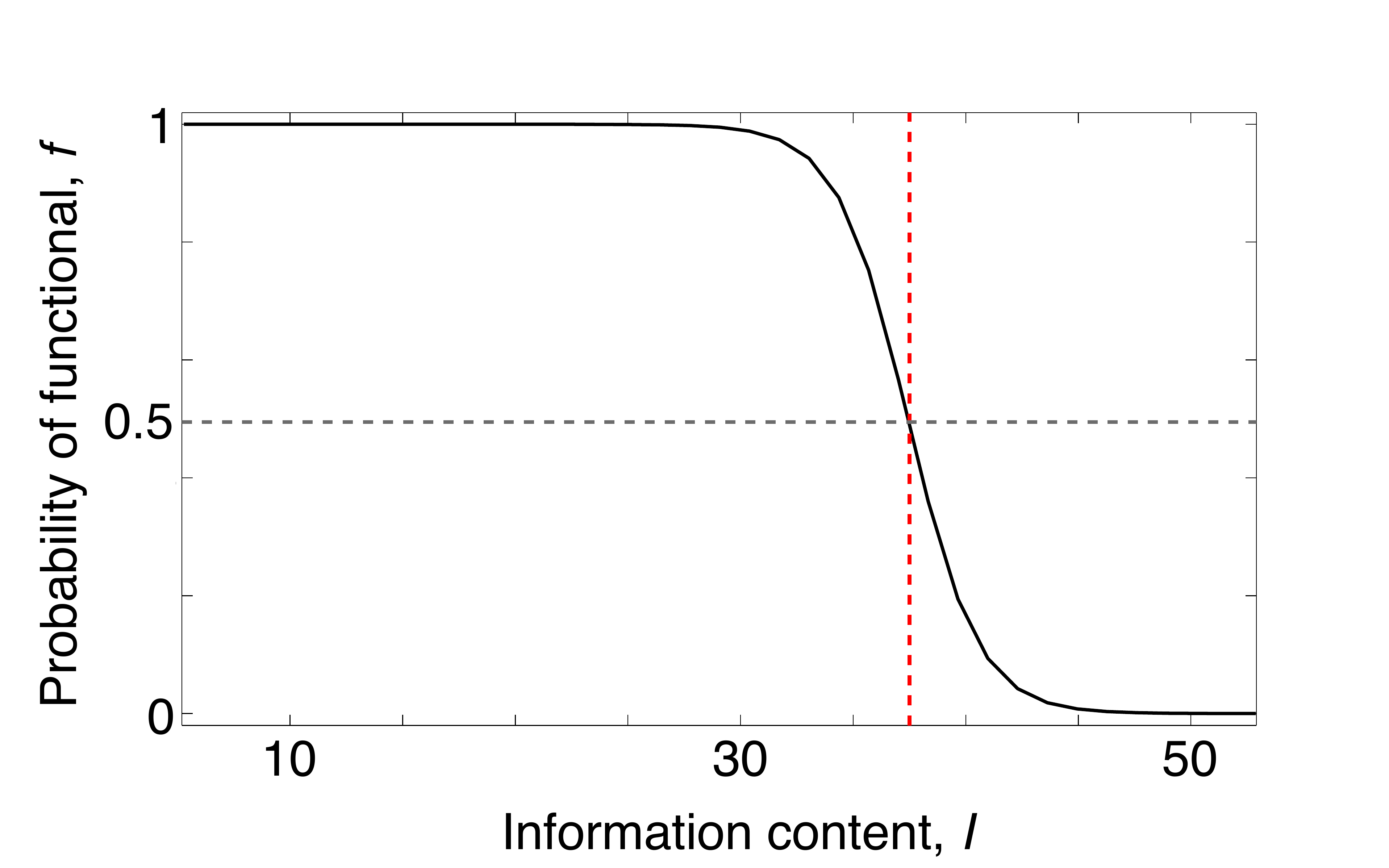}}  \caption*{Figure 1 -- The
equilibrium distribution for the number of mismatched nucleotides at a binding
site under stabilizing selection is bimodal. Top  -- The equilibrium distribution
for a binding site of length $n=20$ with redundancy $r=1$ (corresponding to
information content $I$ = 40 bits) and selection strength $Ns=10$. The dashed line
shows the analytic approximation for the equilibrium distribution, and the points
indicate the results of $10^3$ Monte-Carlo simulations of evolution in the weak
mutation limit (see SI). The left peak in the distribution, centred around
a small number of mis-matched nucleotides, is the result of selection; whereas the
right peak, centred around a large number of mis-matched nucleotides, is caused
by genetic drift.  The bimodal equilibrium distribution suggests a natural
definition for the probability, $f$, that a binding site is functional (Eq. 2
main text): the summed probability of falling near the selected peak.  The bottom
panel shows how the probability of being functional, $f$, depends on the
information content of the binding site, for given selection strength $Ns=10$.
Increasing the information content of the binding site results in a threshold-like
decline in the probability that the binding site is functional.  The gray line
corresponds to functional probability $f=0.5$, and the red line indicates the
information content of the binding site for which this occurs, $I=37.5$ bits in this case.}
\end{figure}

\subsection*{A definition of functional binding}
The bimodal form for the equilibrium distribution of mis-matches is important because it
provides a natural way to separate binding sites into ``functional'' and
``non-functional'', and thus to simplify the  analysis of modules with multiple
binding sites.  
%Separating functional from non-functional sites is not
%straightforward \emph{a priori}, especially because there are many known instances
%in which transcription factor binding sites known to be ``functional'' nonetheless
%have intermediate binding strength.  
We adopt a simple, operational definition for
functional binding sites: all binding sites for which the probability of binding
exceeds one half, i.e. $\pi_i>0.5$, are defined as functional; and all others are
defined as non-functional. This is a natural definition 
because the probability of binding, $\pi_i$, is a sigmoidal function of the
number of mis-matches, $i$, with a fairly sharp threshold occurring at the value
of $i$ for which $\pi_i=0.5$ (the threshold value is given by
$i=\frac{\log[P]}{\epsilon}$).

%DEFER THIS TEXT AND NOTE TO FIGURE ONE CAPTION:
%The definition of a functional binding site makes further sense because the
%equilibrium distribution for the number of mismatches, $F_i$, has one peak on
%either side of this threshold value: one peak occurs at some
%$i<\frac{\log[P]}{\epsilon}$ and the other, neutral peak occurs at some
%$i>\frac{\log[P]}{\epsilon}$ (see Fig.~1). 

Given this definition, the equilibrium
probability, $f$, that a binding site will be functional is given by 
\begin{equation}
f=\sum_{\{i|\pi_i>0.5\}}F_i.
\end{equation}
\\
Fig.~S1 shows how the probability a binding site is functional, $f$, depends upon
the scaled strength of selection, $Ns$.  This relationship displays a sharp,
threshold behavior, as reported in \cite{Gerland:2002fk}.  Thus, as selection
strength or population size increases, binding sites rapidly switch from having
many mismatched nucleotides to having few mismatched nucleotides, so that the
chance of binding rapidly switches from below one-half to above one-half.

%Finally, Fig.~1c shows a contour map for the probability that a binding site is functional, for different values of information content $I$ and
%selection strength $Ns$.  As the figure shows, there are a large range of values of selection strength and information content, relevant to
%observed transcription factor binding sites, for which stabilizing selection cannot maintain functional binding sites.
\subsection*{Evolution of a single binding site}

Before considering complex regulatory modules, we will first analyze the evolution
of a regulatory module composed of a single binding site.  We consider a single
binding site with information content $I$ evolving in a population of size $N$
under stabilizing selection of strength $s$ for binding. As described above, we can 
determine
the probability, $f$ that it is functional in equilibrium.  The information
content of the binding site depends on its length $n$ and average degeneracy $r$,
which are independent parameters.  However, any pair of values $\{n,r\}$ that
result in a given information content, $I=n\log_2\left[\frac{4}{r}\right]$, result
in the same (or very close to the same) probability $f$ of the site being functional
(see SI -- Fig.~S2).  Thus we
confine ourselves to discussing the information content of sites in what follows.

Typically, the strength of selection on transcription factor binding sites is of
order $Ns\sim10$ \cite{Mustonen:2008zr,He:2011ly}. Assuming $Ns=10$, Fig.~1b shows
how the probability the site is functional, $f$, depends on its information
content, $I$.  The figure also indicates the critical value of information content
that results in the binding site being functional with probability one-half; this
occurs when $I=37.5$ bits.  For values of information content greater than this,
the probability that the site is functional declines rapidly to zero;
whereas a binding site with less than this amount of information has probability
of being functional near 1.  Therefore, a regulatory module consisting of a
single binding site will likely be functional whenever $I\leq37.5$, given
selection of strength $Ns=10$.  This simple case forms a basis for comparison as
we consider modules with multiple binding sites, below.

\subsection*{Evolution of regulatory modules composed of two binding sites}

Next, we used our population-genetic model to study the evolution of regulatory
modules composed of two binding sites.
%, before generalising to modules composed of an arbitrary number of binding sites
%executing arbitrary logic.
A pair of binding sites for two co-expressed transcription factors can be bound in
four possible combinations.  We assume that the regulated target gene produces
fitness $1$ if the two sites are bound in a ``desired" combination, and it
produces fitness $1-s$ otherwise.  The desired combination (or combinations) of
transcription factor binding depend on the signal integration task to be executed
by the regulatory module.  For example, if the module carries out an AND logic,
then the fitness of the regulated gene is $1$ when both binding sites are bound
and $1-s$ otherwise.  If the module carries out an OR logic, then the fitness is
$1$ when either one or both of the binding sites is bound.  If the module carries
out an XOR logic, then the fitness is $1$ when one but not both binding sites are
bound.

For a pair of binding sites, $A$ and $B$, with $i$ and $j$ mismatched nucleotides
respectively, the probability of desired binding for an AND regulatory module is
given by $\sigma_{ij}^{AND}=\pi^A_i\pi^B_j$.  For an OR regulatory module the
probability of desired binding is
$\sigma_{ij}^{OR}=\pi_i^A+\pi_j^B-\pi_i^A\pi_j^B$, and for an XOR  regulatory
module the probability of desired binding is
$\sigma_{ij}^{XOR}=\pi_i^A+\pi_j^B-2\pi_i^A\pi_j^B$.  These three possible logics,
and their associated binding probabilities are summarized in Table 1. In this
table cooperativity between the two transcription
factor proteins is neglected. The impact of cooperativity is considered in the SI
(Fig.~S5-S7).

\begin{table*}[h!]
 \caption{Regulatory logics for modules with two binding sites}\begin{center}
 \begin{tabular}{@{\vrule height 10.5pt depth4pt  width0pt} p{2.5cm} p{9.5cm} p{5.5cm}}
 \cellcolor[rgb]{1,1,1}   \sf \textbf{Logic gate}& \sf  \cellcolor[rgb]{1,1,1}
 \textbf{Selected regulation} & \cellcolor[rgb]{1,1,1} \sf \textbf{Probability of
 selected binding} \\
\cellcolor[rgb]{0.9,0.9,0.9}\sf AND &\cellcolor[rgb]{0.9,0.9,0.9} \sf A and B must both be bound for correct regulation&\cellcolor[rgb]{0.9,0.9,0.9} $\pi_{i}^A\pi_{j}^B$\\  
\cellcolor[rgb]{0.8,0.8,0.8} \sf OR &\cellcolor[rgb]{0.8,0.8,0.8} \sf Either A or B or both must be bound for correct regulation&\cellcolor[rgb]{0.8,0.8,0.8} $\pi_i^A+\pi_j^B-\pi_{i}^A\pi_{j}^B$\\  
\cellcolor[rgb]{0.9,0.9,0.9} \sf XOR &\cellcolor[rgb]{0.9,0.9,0.9} \sf Either A or B but not both must be bound for correct regulation&\cellcolor[rgb]{0.9,0.9,0.9} $\pi_i^A+\pi_j^B-2\pi_{i}^A\pi_{j}^B$\\  
%\cellcolor[rgb]{0.8,0.8,0.8} \sf No interaction &\cellcolor[rgb]{0.8,0.8,0.8} \sf A and B contribute independently to correct regulation&\cellcolor[rgb]{0.8,0.8,0.8} $\pi_{ij}^A+\pi_{ij}^B$\\  
\end{tabular} \end{center}
\end{table*}

The equilibrium distribution of mismatched nucleotides $(i,j)$ for the Markov
processes describing the evolution of a two-site regulatory module, in the weak
mutation limit, can be derived analytically (see SI).
Fig.~2 shows contour plots for the probabilities, $f_A$ and $f_B$, that each of
the binding sites are functional in equilibrium, for different values of
information content at each binding site.  The figure shows these contours for
each of the three possible regulatory logics described above, with selective
strength $Ns=10$.
%Also shown for comparison is a plot in the case when the transcription factors
%$A$ and $B$ are expressed mutually exclusively -- i.e. when both binding sites
%form independent modules of size one. EXPLAIN HERE!! HOW DERIVE?  
Also shown for comparison in Fig.~2a is a contour map for two
binding sites functioning independently in isolation of each other.  This
corresponds to the case of a single binding site, as discussed in the previous
section, with the modification that the information content of two such isolated
binding sites, $A$ and $B$, are shown, so that the contour map for this case may
be directly compared with the contour maps for two-site regulatory modules
(Fig.~2b-d).

Fig.~2 illustrates the central result of our study: 
for all three possible regulatory logics,
the range of information content for which both binding sites are functional with
high probability is much smaller, and occurs for lower information content, than
the same region for a regulatory module consisting of a single binding site.
Thus, our analysis predicts that functional binding sites belonging to two-site
modules, regardless of their regulatory logic, will tend to have less information
than binding sites occurring in single-site modules.

\begin{figure}[h!]
\centering
%\subfigure{\includegraphics[scale=0.5]{IndN1}} 
%\subfigure{\includegraphics[scale=0.5]{andN1}} 
%\subfigure{\includegraphics[scale=0.5]{xorN1}} 
%\subfigure{\includegraphics[scale=0.5]{orN1}} 
\includegraphics[scale=0.25]{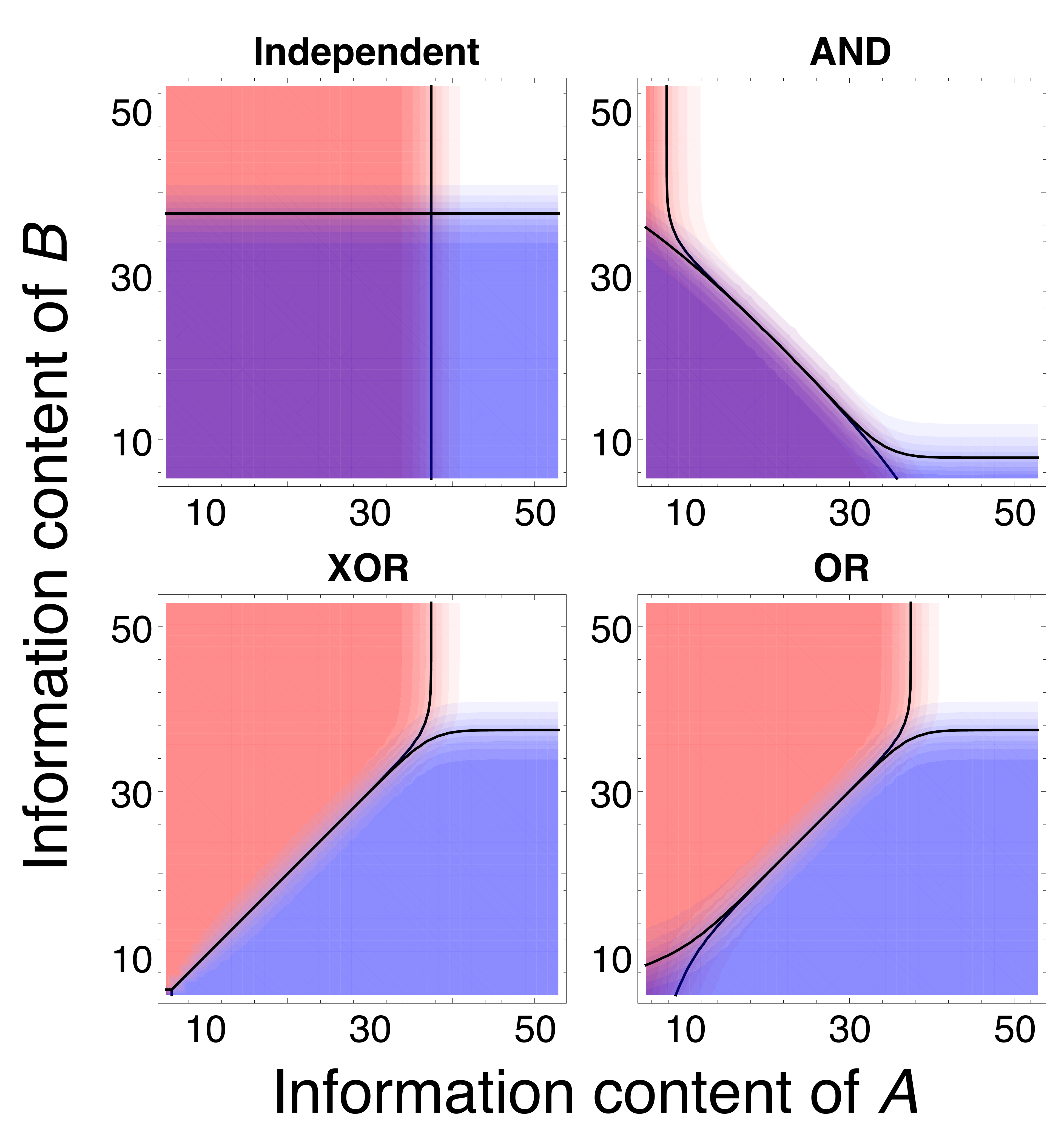}
\caption*{Figure 2 -- Regulatory modules containing two binding sites are composed of 
individually less specific
binding sites, compared to a module composed of a single binding site.
Each panel shows a pair of overlapping contour plots with the
probability that binding site $A$ is functional, $f_A$, in red and the probability
that binding site $B$ is functional, $f_B$, in blue. Black lines indicate the
contours $f_A=0.5$ and $f_B=0.5$. Purple regions indicate that both binding sites are
likely to be functional, red regions indicate that only $A$ is likely to be
functional, blue regions indicate that only $B$ is likely to be functional, and
white regions indicate that neither site is likely to be functional. 
%The x- and
%y-axes give the information content for the binding motifs of $A$ and $B$. 
All
plots are generated with selection strength $Ns=10$.  Clockwise from top left, the figures
show: (i) Probability of being functional for individual binding sites in isolation: The purple
region, which occurs when both binding sites have $I\lesssim37.5$ bits, serves as a
basis for comparison with two-site binding modules. (ii) A two-site module with AND logic, 
so that both $A$ and $B$ are selected to
be bound. The 
purple region is smaller than in the one-site case, indicating that functional binding 
sites maintained in the module
will contain less information than in the one-site case. (iii) A two-site module
with OR logic, so that $A$ or $B$ is selected to be bound. Only a small
purple region at low information content is visible. As a result, only the binding
site with lower information content will typically remain functional.
(iv) A
two-site module with XOR logic,
so that $A$ or $B$ but not both are selected to be bound; again, the binding sites maintained
in such a two-site module each have less information than in a single-site module.} \end{figure}

\subsection*{Evolution of regulatory modules composed of many binding sites}

Our analysis of two-site regulatory modules can be naturally extended to describe
larger regulatory modules.  We consider a set of $M$ co-expressed transcription
factors whose binding sites co-regulate a target gene.  The group of $M$ binding
sites can be bound by their respective transcription factors in $2^M$ possible
combinations.  Each combination of bound sites can, in turn, constitute a
desirable or an undesirable pattern of gene regulation.  As a result there are
$2^{2^M}$ possible logics that can be executed by a regulatory module composed of
$M$ binding sites.  An example is illustrated in Table S1 (see SI).  Analogously to the
two-site case, we can construct a function $\sigma_{i_1,i_2,.....i_M}$ to describe
the probability that a set of transcription factor binding sites,
$\left\{A_1,A_2,.....A_M\right\}$, with $\left\{i_1,i_2,.....i_M\right\}$
mismatched nucleotides is bound in a desirable pattern, for a given regulatory
logic.  The equilibrium distribution of the Markov process for the number of
mismatched nucleotides at each binding site can once again be found analytically
in the weak-mutation limit.  This expression is given in the Supporting
Information, however the combinatorial explosion in the number of possible
regulatory logics with module size means that a detailed analytical exploration of
modules with more than $2$ binding sites quickly becomes impractical.

Instead we performed evolutionary simulations in the weak mutation limit.  We
constructed regulatory modules in our simulations as follows: for each binding
site we drew a energy contribution per mismatch, $\epsilon$, and a number of
proteins per cell, $P$, from a uniform distribution in the empirically determined
ranges $1\leq \epsilon \leq 3$ and $10^0\leq P\leq 10^3$, respectively. We also
drew a binding site length, $n$, and average degeneracy, $r$, from a uniform
distribution in the range $5\leq n\leq40$ and $1\leq r<4$ respectively, with the
additional constraint that we condition on
$n\left(1-\frac{r}{4}\right)>\frac{\log[P]}{\epsilon}$, in order to ensure that
selection is able to differentiate between functional and non-functional sites.

We can easily extend our analysis of two-site modules by assuming a strict AND
logic, or an OR logic, across all the binding sites in the multi-site module. In
addition, we explored random logics as well. To do this, we chose a logic by
selecting uniformly from among the $2^{2^M}$ possible logics.

Each module was then allowed to evolve until $>10^2$ mutations fixed, in order to
ensure that equilibrium has been reached, and the simulation was then stopped.
Once the simulation was stopped, a module is defined as functional if each of its
composite binding sites is functional.  We calculate the average information
content for all the functional modules of a given size.  Because the parameter
space we are sampling is large we constructed $>10^8$ different regulatory modules
of each size, with each module composed of different binding sites. Of these
modules, typically $<10^2$ are found to be functional. 

Fig.~3 shows the relationship between the average information content of binding
sites, and module size.  We performed simulations using the randomly generated
mixed logics, described above, as well as modules composed of randomly chosen binding
sites executing AND and OR logics.  In all three cases we observe a decline in the
average information content per binding site as module size $M$ increases.  In
addition, we observe a decline in the ensemble variance in information content of binding
sites with module size.  Thus, the region of parameter space in
which functional modules are maintained by selection becomes smaller and it includes
lower-information binding sites as the module size, and hence regulatory
complexity, increases.

\begin{figure*}[h!]
\centering
\includegraphics[scale=0.15]{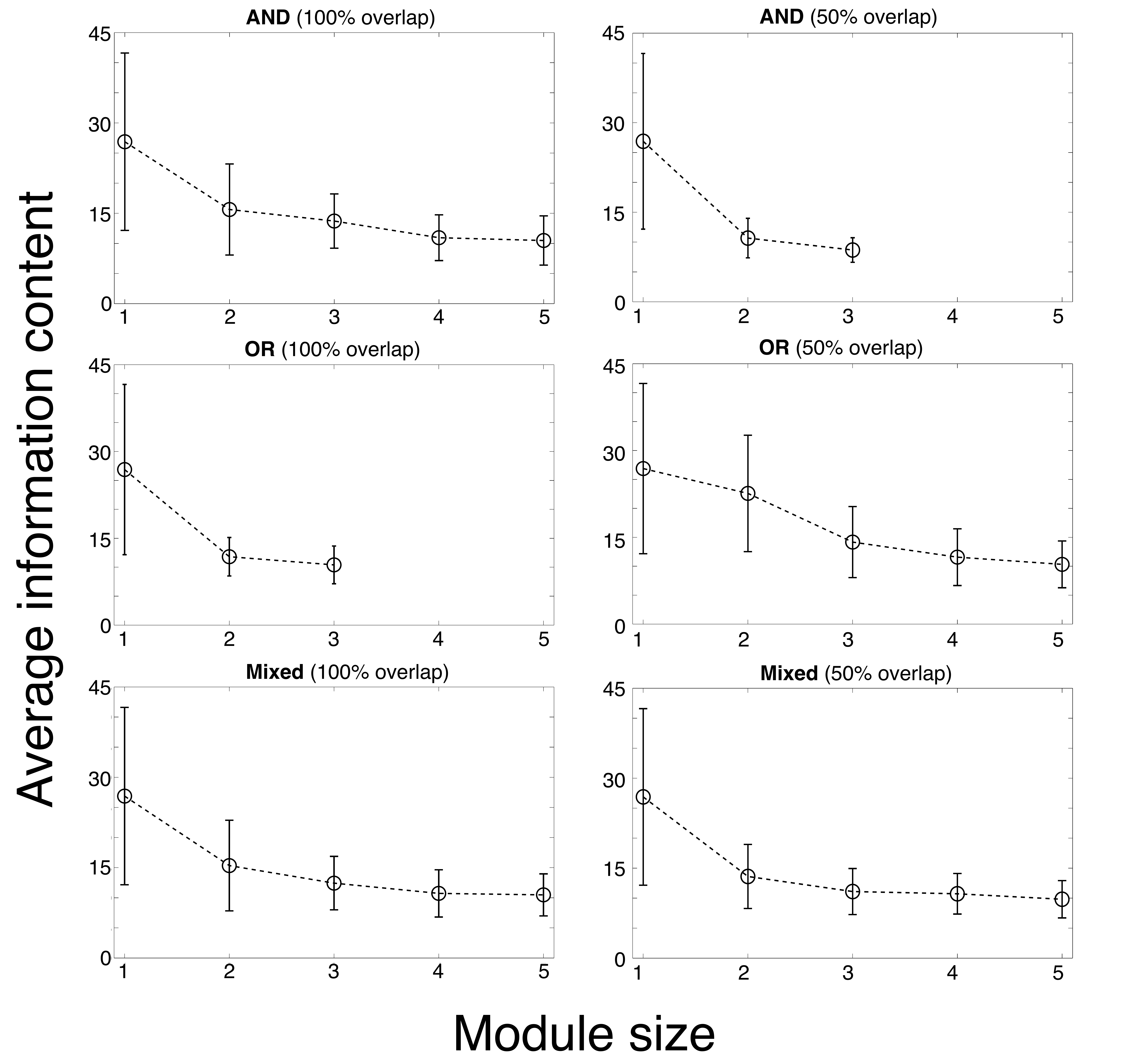}
%\subfigure{\includegraphics[scale=0.25]{NewAND100}} 
%\subfigure{\includegraphics[scale=0.25]{NewAND50}} 
%\subfigure{\includegraphics[scale=0.25]{NewOR100}} 
%\subfigure{\includegraphics[scale=0.25]{NewOR50}} 
%\subfigure{\includegraphics[scale=0.25]{NewMIX100}} 
%\subfigure{\includegraphics[scale=0.25]{NewMIX50}} 
\caption*{Figure 3 -- Information content of binding sites is predicted to
decrease with module
size, regardless of regulatory logic. Points show the ensemble average information 
content per binding site in a module, and bars show the ensemble standard deviation 
(bar width is 2SD either side of the mean). Panels top to
bottom show modules with AND, OR and Mixed (arbitrary) logics. 
The left-hand column (100\% overlap) corresponds to a model in which all
all transcription factors are co-expressed, at all times. The right-hand column
(50\% overlap) corresponds to a model in which any given pair of factors are co-expressed
half the time. Monte Carlo simulations of binding site 
evolution in the weak selection limit were performed, as described in the main text.
For each module size, replicate simulations were performed until at least $10^2$
functional modules were produced. (Missing data points indicate that no functional
modules were produced, after even $10^6$ simulations.) All
modules were evolved with selection strength $Ns=10$. 
In all cases, the average information content of the functional binding
sites in a module, and the ensemble variance of information content among functional
binding sites, decrease with module size, $M$.
} \end{figure*}

\subsection*{The environmental schedule of co-regulating transcription factors}

So far we have focused on regulatory modules whose participating transcription
factors are always co-expressed.  Generally, of course, this is not the case, as
transcription factors themselves are up-regulated and down-regulated in different
patterns as the environment changes \cite{Tirosh:2008kx}.  To account for this we modified our model to
include changes in the patterns of transcription factor expression as cells
experience different environmental conditions.  We assumed that in a given
condition a transcription factor is either fully expressed (i.e ON) or not
expressed at all (i.e OFF).  Whether or not a target gene is properly regulated
depends as before on whether or not its transcription factor binding sites are
bound in a desirable pattern.  This in turn depends both on the pattern in which
regulating transcription factors are expressed, and on the number of mismatched
nucleotide at each transcription factor binding site.

This analysis is best understood in the case of a two-site module.  For a pair of
transcription factors $A$ and $B$, we analyzed environmental variation such that
transcription factor $A$ is expressed without factor $B$ in a proportion $\tau_A$
of environmental conditions; and likewise factor $B$ is expressed without $A$ in
a proportion $\tau_B$ of conditions; and, finally, both transcription factors are
co-expressed in a proportion $\tau_{AB}$ of conditions, with
$\tau_A+\tau_B+\tau_{AB}=1$.  The function $\sigma_{ij}$ for the probability that
a two-site regulatory module is bound in a desired configuration now depends
\emph{both} on the probability that each transcription factor is expressed and the
probability that they are bound to their binding site.  The resulting equilibrium
behavior is summarized in Table 2.

\begin{table*}[h!] 
 \caption{Regulatory logics for two-site binding modules, when transcription factors are not always co-expressed}\begin{center} \begin{tabular}{@{\vrule height 10.5pt depth4pt  width0pt} p{0.8cm} p{3cm} p{3cm} p{3cm} p{6.5cm}}\ \
 &\textcolor[rgb]{0,0,0}{\sf \textbf{Both expressed}} &\textcolor[rgb]{0,0,0}{\sf \textbf{Only A expressed}} &\textcolor[rgb]{0,0,0}{\sf \textbf{Only B expressed}} &\ \sf \textbf{Probability of selected binding} \\
\sf \textbf{AND} &\cellcolor[rgb]{0.9,0.9,0.9}\ \ \textcolor[rgb]{0.0,0.0,0.0}{\sf A AND B} &\cellcolor[rgb]{0.9,0.9,0.9}\ \ \ \ \ \ \ \textcolor[rgb]{0,0,0}{\sf  --} &\cellcolor[rgb]{0.9,0.9,0.9}\ \ \ \ \ \ \ \textcolor[rgb]{0,0,0}{\sf --} &\cellcolor[rgb]{0.9,0.9,0.9} $\tau_{AB}\pi_{i}^A\pi_{j}^B$\\  
\ \ \sf \textbf{OR} &\cellcolor[rgb]{0.8,0.8,0.8}\ \ \ \textcolor[rgb]{0.0,0.0,0.0}{\sf A OR B} &\cellcolor[rgb]{0.8,0.8,0.8}\ \ \ \ \ \ \ \textcolor[rgb]{0,0,0}{\sf A} &\cellcolor[rgb]{0.8,0.8,0.8}\ \ \ \ \ \ \ \textcolor[rgb]{0,0,0}{\sf B} &\cellcolor[rgb]{0.8,0.8,0.8}$\tau_A\pi_i^A+\tau_B\pi_j^B+\tau_{AB}(\pi_i^A+\pi_j^B-\pi_{i}^A\pi_{j}^B)$\\  
\sf \textbf{XOR} &\cellcolor[rgb]{0.9,0.9,0.9}\ \ \textcolor[rgb]{0.0,0.0,0.0}{\sf A XOR B}&\cellcolor[rgb]{0.9,0.9,0.9}\ \ \ \ \ \ \ \textcolor[rgb]{0,0,0}{\sf A} &\cellcolor[rgb]{0.9,0.9,0.9}\ \ \ \ \ \ \ \textcolor[rgb]{0,0,0}{\sf B} &\cellcolor[rgb]{0.9,0.9,0.9}$\tau_A\pi_i^A+\tau_B\pi_j^B+\tau_{AB}(\pi_i^A+\pi_j^B-2\pi_{i}^A\pi_{j}^B)$\\
\end{tabular} \end{center}
\end{table*}

What is the impact of varying the environmental schedule on our
results? If we set $\tau_{A}=\tau_{B}=\frac{1}{2}\left(1-\tau_{AB}\right)$ then we
can simply consider variation in $\tau_{AB}$, i.e the proportion of time that both
$A$ and $B$ are co-expressed. When $\tau_{AB}=1$ both factors are always
co-expressed, as in our analysis above, and the probability that their binding sites 
are functional varies
with information content shown in Fig.~2. As $\tau_{AB}$
decreases, so that transcription factors are not always expressed at the same time,
we still find that complex regulation requires binding sites with
less information. The precise range of information contents that result in both binding sites
being functional changes in a way that depends on the logic of the module.  For OR
and XOR logics, the range of information contents that results in functional
binding for both sites increases  as $\tau_{AB}$ decreases (Fig.~S9-S11), and so the average information
content of the pair of sites will tend to increase. However for AND logic, as
$\tau_{AB}$ decreases the range of information contents that result in both sites
being functional decreases.  When $\tau_{AB}=0$, $A$ and $B$ are always expressed
independently. For OR and XOR logic this means that the probability that both
binding sites are functional varies with the information content of the sites  in
the same way as for a single site module (Fig.~2). For AND logic, when
$\tau_{AB}=0$, it is not possible to maintain both binding sites, because the transcription factors are never co-expressed.

%We find that, in the case of the OR and XOR regulatory modules, the occurrence of environmental conditions in which $A$ or $B$ are expressed in isolation tends to increase the ability of selection to maintain the binding sites of both transcription factors (see Fig.~SXXXX).
%In fact, when $\tau_{AB}=0$, each binding site is selected on independently, and we return to the case shown in Fig.~2a.

%WHAT ELSE DO WE SAY ABOUT THIS CASE? RE-ITERATE MAIN RESULT!l? EXPLAIN THAT AS EVEN WHEN TAB>0, BOTH FUNCITONAL BINDING SITES
%HAVE LESS INFORMATION THAN A SINGE-SITE MODULE. BUT AS TAB->0, WE RECOVER THE CASE OF HIGH_INFORMATION SINGE-SITE MODULE, AS EXPECTED. COMBINE THIS
%WITH PARAGRPH ABOVE.

Our analysis of variation in the co-expression of transcription factors naturally
extends to arbitrary modules of $M$ binding sites.  To study this in the simplest case, we
assumed that each transcription factor is expressed with fixed probability, independent
of the environment.  We then varied the expression overlap of all
pairs of transcription factors by varying this probability.  As shown in Fig.~3,
decreasing the expression overlap of factors participating in a module alters how
quickly the average information content of binding sites in functional modules
declines with module size.  For example, functional modules executing AND logics
experience a more rapid decline in the information content of their composite
binding sites, and functional modules executing OR logic experience a slower
decline.  Nonetheless the general trend we have discovered continues to hold: the mean
and the ensemble variance of the information content of a binding site in a module 
declines with module size, even when transcription factors are not always
expressed at the same time.

\subsection*{Empirical data on site specificity and module size}

We have demonstrated that as the size and complexity of regulatory modules
increase, the average and ensemble variance in the information content of the binding sites
maintained by selection will decrease.  How can we  assess the
complexity of regulatory modules to compare these predictions to empirical data?  
We have argued that the
simplest way to do this is to look at the number binding sites regulating a target
gene.  Such an argument is supported by a comparison of prokaryotes and
eukaryotes, where the regulatory complexity of eukaryotes tends to be greater
\cite{Lynch:2006vn,Janga:2008ys}.  Accordingly we find that, in a comparison of
{\it Saccharomyces
cerevisia} and \textit{Escherichia coli}, the yeast genes have more regulators, and yeast
transcription factors have lower information content binding sites.  
However, such a comparison is insufficient on its own to confirm the
predictions of our model because of the well known and often discussed differences
in the mechanisms of transcription regulation used by prokaryotes and eukaryotes
\cite{Struhl:1999ly,Raj:2008bs}.

A better way to empirically test our prediction is to consider variation in
regulatory complexity within a given species.  For this purpose we used the yeast
transcription network  \cite{ABC:Lee2002,ABC:Har2004,Bryne:2008fk}, which benefits from extensive study and available datasets.
We calculated the correlation between the number of binding sites that regulate a
given target gene, and the average information content of the associated binding motifs.
Here we found, as predicted by our analysis, a strong negative correlation between
the number of regulators of a gene and the average information content of the
regulatory motifs (Fig.~4).  As predicted by our analysis, we also found a negative
correlation between the number of regulators of a gene and the variance in the
information content of its regulatory motifs across targets (Fig.~4).

\begin{figure}[h!] \centering
%\subfigure{\includegraphics[scale=0.25]{ExpTG}}
%\subfigure{\includegraphics[scale=0.27]{BitTG}}
%\subfigure{\includegraphics[scale=0.27]{InfVar}}  
\includegraphics[scale=0.25]{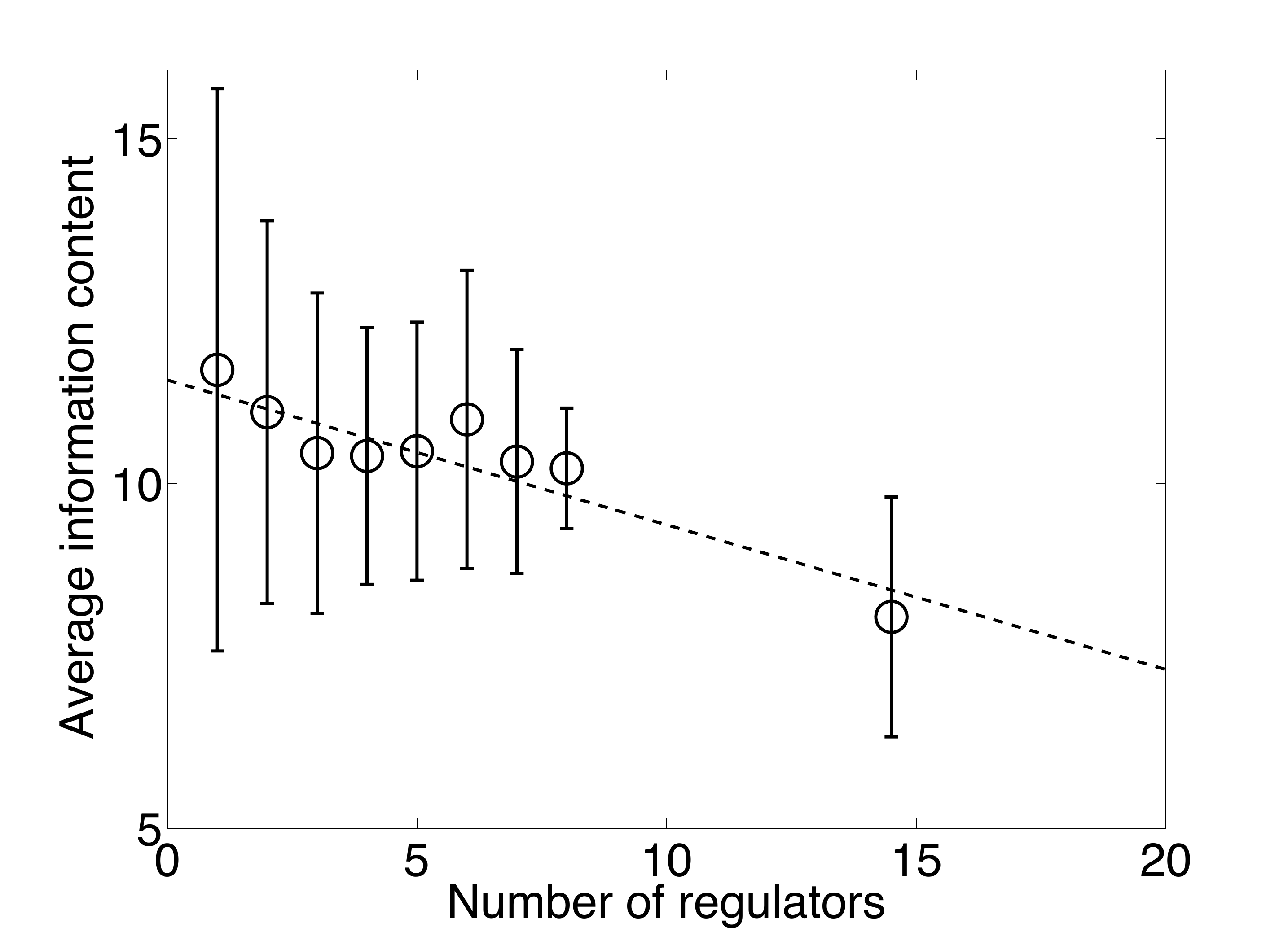} \caption*{Figure 4 -- The empirical
relationship between the number of regulators of a gene and the information
content of its regulatory binding sites, in the yeast transcription network
\cite{ABC:Lee2002,ABC:Har2004,Bryne:2008fk}. Both the average information content
of regulatory motifs (points) and its standard deviation (bars) decrease with the
number of regulators.  The dashed line is a linear fit to the data and shows a
significantly negative slope ($p<2\times10^{-8}$).  Points show the average
information content of binding motifs, for target genes binned according to the
number of regulators. Bin sizes were chosen so that each bin contains at least 10
target genes. Bars show the standard deviation in the information content of
binding sites, taken across all genes with a given number of regulators. Bars
extend 1SD either side of the mean. There is a negative relationship between the
standard deviation information content and the number of regulators
($p<1\times10^{-4}$).  For both the mean and ensemble standard deviation
information content, the negative relationship observed in the yeast data
coincides with the predictions of our population-genetic model.  } \end{figure}

Similar correlations to these, also within the yeast transcription network, were
reported previously by Bilu and Barkai \cite{Bilu:2005kx}.  They found an increase
in the ``fuzziness'' at a given transcription factor's binding sites with
the number of other regulators at a gene. This observation can also be understood
within the context of our evolutionary model, 
%where we find that the number of mismatches at binding sites with a 
because binding site information content decreases with module
size, and lower information content binding sites tend to have greater degeneracy per nucleotide. 
Bilu and Barkai also observed a correlation between the variance in the
expression of genes across environmental conditions, and the number of
transcription factors regulating them.  We observed the same correlation.
This observation supports our interpretation that more regulators indicates
greater regulatory complexity: genes with greater expression variation across
environmental conditions require more complex regulation in order to perform the
more complex signal integration tasks required for them to respond appropriately
to different environments.

Bilu and Barkai suggested that their observations can be explained either by
weaker selection on the expression of genes with highly variable expression across
conditions, or by a tendency for multiple co-regulating binding sites to
experience compensatory mutations leading to greater fuzziness.  Our analysis does
not rule out these explanations.  However we have shown that no such assumptions
are necessary to explain their observations: large, complicated regulatory modules
are \textit{a priori} expected to contain low-information binding sites, even
under a constant strength of selection.

\section*{Discussion and Conclusion}

%[SUMMARIZE RESULTS IN TWO SENETENCES, FIRST PARAGRAPH]

Using a standard population-genetic model, we have shown that as the size of a
regulatory module increases, the specificity of its constituent binding sites must
tend to decrease if they are to be maintained.  This general result
does not depend on the specific logic being executed by a module, nor does it
depend on the pattern of expression of the transcription factors that participate
in the module.  

Our results do not tell us how ``best'' to construct a
regulatory module for a given task, a question which has been addressed elsewhere
\cite{Buchler:2003uq,Little:2010zr,ABC:Lyn2007}.  Rather, our results hold regardless of the logic favored
by selection.  In a certain light our results make simple, intuitive sense: it is
\emph{all} of the nucleotide positions that belong to a regulatory module that
form the target of selection, and so as module size increases, the maximum
mutation rate at any given site, and hence its maximum information content, must
decrease.  What is not obvious, however, is that this effect should be so
pronounced, given the information content and selection strengths typical of
transcription factor binding sites.  Yet this is precisely what we find.

We have argued that our results help explain the observed variation in the
specificity of regulators at different genes within the yeast transcription
network and that they also help explain the tendency for eukaryotes to have lower
information content binding sites than prokaryotes.  On the basis of this, we
would also expect that higher eukaryotes should have lower-information content
binding sites than yeast, due to the greater regulatory complexity required to
execute, for example, tissue-specific gene expression.  A comparison of mouse and
yeast binding sites supports this expectation
\cite{Bryne:2008fk,ABC:Lee2002,ABC:Har2004,Gama-Castro:2011kx}, with mouse tending
to use lower information content binding sites than yeast (\emph{E. coli} $=14.9$
bits, yeast $=12.1$ bits, mouse $=10.6$ bits).  Although the population genetics
of higher eukaryotes are complicated by the presence of frequent recombination,
this is not likely to have an important impact on our study, since we are
concerned with the evolution of only a small number of nucleotides at a particular
genomic position. 

Our analysis was conducted in the limit of weak mutation.  As a result it is
important to consider the timescales over which the evolutionary changes being
described occur.  Because transcription factor binding sites are short ($\sim10$
nt) the system typically reaches equilibrium in $<100$ selective sweeps.  However
the rate of selective sweeps is low, occurring once every $10^{6}$--$10^{8}$
generations.  In yeast, which has a rapid cell division time ($\sim2$hrs), this
means that regulatory modules composed of binding sites that are ``too specific''
under our model will be lost under stabilizing selection over the course of tens
of thousands of years.  This time is short enough to produce the genome-wide
association between the number of regulators and binding-site specificity
according to the processes we have described.  Similarly, time scales are short
enough to be relevant to the differences in binding site information content among
species.
%However the timescales involved imply that the loss of high-information binding
%sites from complex regulatory modules under stabilising selection, will only
%occur as a general trend across genomes [I DONT UNDERSTAND OR CANNOT PARSE THE
%GRAMMER OF THIS SENTENCE].

In constructing our model we have assumed that transcription factors bind to their
sites independently when participating in a module.  In reality, factors in the
same regulatory module may sometimes bind cooperatively \cite{Davidson:2002fk,Buchler:2003uq,Li:2010ys}.  To
determine how such interactions between factors alter our results we analysed the
case of two-site modules with cooperative binding.  The results are shown in the
SI. Fig.~S5-S7.  For AND, OR and XOR logic gates, interactions among factors alter
the shape and size of the region in which both sites are maintained under
stabilizing selection.  However, this region is always limited to lower
information-content sites, compared to the case in which sites are not part of a
module.  We also considered the case in which the increase in binding energy with
each mismatched nucleotide, $\epsilon$, decreases with information content.  This
has the effect that higher-information content sites can suffer a greater number
of mismatches before becoming non-functional, compared to lower-information
content sites.  We find (see SI, Fig.~S8) that although this correlation does
allow higher-information content binding sites to be maintained on average, the
negative correlation between module size and binding site information content
persists.  Finally, we considered the impact of changing selection strength on our
results.  When selection is very strong ($Ns=100$), our results do not hold, since
all binding sites are maintained, regardless of their information content.
Similarly, when selection is very weak ($Ns=1$), virtually no binding sites are
maintained regardless of information content.  However for intermediate values
of selection strength ($1<Ns<20$) our results hold (see SI, Fig.~S3-S4), and these
are the values of selection strength typically associated with transcription
factor binding sites \cite{Mustonen:2008zr,He:2011ly}.

%I EXPECT YOU'LL HATE THIS BUT.....

Higher eukaryotes must carefully orchestrate gene expression in order to produce
the elaborate phenotypes associated with multi-cellularity. And even simple
eukaryotes require complex regulation of genes that must respond to different
environmental conditions. In spite of this, many eukaryotic genes have noisy
expression, and many of the transcription factors that regulate them bind weakly.
Our study suggests an evolutionary perspective on this phenomenon: complexity
requires some sloppiness.

\section*{Acknowledgement}

\end{document}